\documentstyle[12pt,aaspp4]{article}

\begin{document} 
\title{The Carina Project: I. Bright Variable Stars\footnote{Based on 
observations collected at the European Southern Observatory, La Silla, 
Chile on Osservatorio Astronomico di Capodimonte guaranteed time.}}

\author{M. Dall'Ora\altaffilmark{1,2,3},
V. Ripepi\altaffilmark{1}, F. Caputo\altaffilmark{2},
V. Castellani\altaffilmark{2,4}, G. Bono\altaffilmark{2},
H. A. Smith\altaffilmark{5}, 
E. Brocato\altaffilmark{6}, R. Buonanno\altaffilmark{2,3},
M. Castellani\altaffilmark{2},
C. E. Corsi\altaffilmark{2}, M. Marconi\altaffilmark{1},
M. Monelli\altaffilmark{2,3}, M. Nonino\altaffilmark{7},
L. Pulone\altaffilmark{2},
A. R. Walker\altaffilmark{8}} 

\affil{1. INAF - Osservatorio Astronomico di Capodimonte, Via
Moiariello 16, 80131 Napoli, Italy; dallora, ripepi@na.astro.it} 
\affil{2. INAF - Osservatorio Astronomico di Roma, Via Frascati 33, 
00040 Monte Porzio Catone, Italy; bono, buonanno, caputo, corsi, dallora, 
mkast, monelli, pulone, vittorio@mporzio.astro.it} 
\affil{3. Universit\`a  Tor Vergata, Via della Ricerca Scientifica 1, 
00133 Rome, Italy}
\affil{4. INFN - Sezione di Ferrara, via Paradiso 12, 44100 Ferrara, Italy.} 
\affil{5. Dept. of Physics and Astronomy, Michigan State University, 
East Lansing, MI 48824, USA; smith@pa.msu.edu}
\affil{6. INAF - Osservatorio Astronomico di Collurania, Via M. Maggini, 
64100 Teramo, Italy; brocato@te.astro.it} 
\affil{7. INAF - Osservatorio Astronomico di Trieste, Via G.B. Tiepolo 11, 
40131 Trieste, Italy; nonino@ts.astro.it}   
\affil{8. Cerro Tololo Inter-American Observatory, National Optical Astronomy
Observatories, Casilla 603, La Serena, Chile; awalker@noao.edu}

\date{Received ; accepted  }

%________________________________________________________________
\clearpage 
\begin{abstract}

We present new BV time series data of the Carina dwarf spheroidal (dSph)
galaxy.  
Current data cover an area of $\approx0.3$ degree$^2$ around the center 
of the galaxy and allowed us to identify 92 variables. Among them 75 
are RR Lyrae stars, 15 are "bona fide" Anomalous Cepheids, one might 
be a Galactic field RR Lyrae, and one is located along the Carina 
Red Giant Branch (RGB). Expanding upon the seminal photographic 
investigation by Saha, Monet, \& Seitzer (1986) we supply 
for the first time accurate estimates 
of their pulsation parameters (periods, amplitudes, mean magnitude and 
colors) on the basis of CCD photometry. Approximately 50\% of both 
RR Lyrae and Anomalous Cepheids are new identifications. Among the 
RR Lyrae sample 6 objects are new candidate double-mode ($RRd$) 
variables. 
On the basis of their pulsation properties we estimated that two variables  
(V152, V182) are about 50\% more massive than typical RR Lyrae stars, while 
the bulk of the Anomalous Cepheids are roughly a factor of two more massive 
than fundamental mode ($RRab$) RR Lyrae stars. This finding supports the 
evidence that these objects are intermediate-mass stars during central He 
burning phases.  

We adopted three different approaches to estimate the Carina distance 
modulus, namely the First Overtone Blue Edge (FOBE) method, the 
Period-Luminosity-Amplitude ($PLA$) relation, and the 
Period-Luminosity-Color (PLC) relation. We found $DM=20.19\pm0.12$, 
a result that agrees quite well with similar estimates based on different 
distance indicators.   

The data for Carina, together with data available in the literature, strongly 
support the conclusion that dSph galaxies can barely be classified into 
the classical Oosterhoff dichotomy. The mean period of $RRab$ in Carina 
resembles that found for Oosterhoff type II clusters, whereas the ratio 
between first overtones ($RRc$) and total number of RR Lyrae is quite 
similar to that found in Oosterhoff type I clusters. 
\end{abstract}

\keywords{galaxies: dwarf -- galaxies: individual: (Carina)
-- galaxies: Local Group -- stars: distances -- stars: horizontal branch
-- stars: oscillations}

%%%%%%%%%%%%%%%%%%%%%%%%%%%%%%%%%%%%%%%%%%%%%%%%%%%%%%%%%%%%%%%%%%%%%%%%%%%%
%				Introduction 
%%%%%%%%%%%%%%%%%%%%%%%%%%%%%%%%%%%%%%%%%%%%%%%%%%%%%%%%%%%%%%%%%%%%%%%%%%%%
\pagebreak 
\section{Introduction}
As suggested in several recent studies, and extensively reviewed by 
Mateo (1998, hereinafter M98) and van den Bergh (1999, hereinafter 
VDB), dSph galaxies are the most common type of galaxies in the Local
Group (LG), and presumably in the present-day Universe. Therefore, they 
are as cornerstones on the path to understanding several aspects of 
galaxy evolution. They are metal-poor systems (M98, VDB), with a metal 
content, $Z<$ 0.001, that resembles the metallicity of Galactic Globular 
Clusters (GGCs). On the other hand, only a few of them, namely Tucana 
(Lavery et al. 1996; Castellani et al. 1996) and possibly Ursa Minor 
(Carrera et al. 2002; Ikuta \& Arimoto 2002) and Draco (Dolphin 2002), 
host a {\it single} stellar population older than 
$\approx 10$ Gyr that might be coeval to GGCs. As a whole, dSphs 
are composite systems which are characterized by a sizable  
intermediate-age (4-8 Gyr) stellar component (VDB 1999),  
and show compelling evidence of multiple star-formation episodes 
that range from $\sim$ 12-14 Gyr to very recent epochs $\sim$
2-3 Gyr. Although it was at first suggested that star formation in 
the Leo I dwarf spheroidal began only $\sim$ 7 Gyr ago, it is now 
known that, like the other dwarf spheroidals, it contains an underlying 
ancient stellar population (Caputo et al. 1999; Gallart et al. 1999; 
Held et al. 2000, 2001). 
Empirical facts above discussed indicate that the old stellar populations 
in LG dSphs are approximately coeval with GGCs (Da Costa 1999).  
At the same time, current data also suggest that these systems 
retained substantial amount of gas for a significant fraction of 
the Hubble time. In fact, current photometric surveys support the 
evidence that dSphs experienced star formation episodes that lasted 
several Gyrs. Moreover and even more importantly, in several 
dSphs the most recent star formation episodes date back a few 
Gyr ago. This circumstantial evidence is at odds with current 
theoretical predictions concerning the formation and evolution 
of LG dSph (see Moore et al. 1999 and Monelli et al. 2002, hereinafter 
Paper II).   

In the dSph realm, the Carina dSph is one of the the most prominent 
examples of multiple stellar populations, since its
Color-Magnitude Diagram (CMD) shows a surprisingly complex
star-formation history. On the basis of the comparison between 
the multiple Main Sequence Turn Off and theoretical isochrones, it
has been suggested that this galaxy underwent significant bursts
of star formation at 3, 7, and 15 Gyr ago (Hurley-Keller, Mateo,
\& Nemec 1998; Smecker-Hane et al. 1996; Paper II). Moreover, Smecker-Hane
et al. (1994) showed that two morphologically distinct Horizontal
Branch (HB) structures are present: the former ("old") has a
visual magnitude $V\sim$ 20.65 mag at the RR Lyrae gap and extends
to bluer colors, the latter ("young") is located close to the RGB 
and it is $\sim$ 0.25 mag brighter than the old one. 
Interestingly, the periods of the RR Lyrae stars in Carina (Saha, 
Monet \& Seitzer 1986; hereinafter SMS) suggest that this galaxy, like 
most dSph galaxies, such as Draco (Baade \& Swope 1961), is a 'bridge' 
between Oosterhoff type I (OoI) and type II (OoII) clusters. 

The Carina Project was driven by these stimulating peculiarities 
and aims at fully exploiting modern technological capabilities 
of wide field imagers to improve current knowledge on the stellar 
content of dSph galaxies in the LG. In this context it is worth 
mentioning that Carina is a fundamental laboratory, since it is 
relatively close and the CMD can allow us to supply a complete census of 
its stellar content on a star-by-star basis from the tip of the 
RGB down to the Turn-Off of the old population. At the same time, 
Carina is also a fundamental laboratory to provide new insights 
on the properties of variable stars such as Anomalous Cepheids
(Bono et al. 1997a), RR Lyrae stars (SMS), and dwarf Cepheids 
(Mateo et al. 1998). Moreover, detailed information on static and 
variable stars will supply tight constraints on the evolutionary 
history of Carina. 

Based on a series of exposures collected with the Wide Field Imager
(WFI) available at the 2.2m ESO/MPI telescope (La Silla, Chile), we
determined $B$ and $V$ magnitudes for $\sim$ 68,000 stars,
extending from the tip of the RGB ($V\sim$ 17.5 mag) down
to $V\sim$ 25.5 mag. Photometric data of the observed static stars
are discussed in a companion paper (Paper II), while the present 
investigation deals with the bright pulsating variables, namely 
RR Lyrae and Anomalous Cepheids.
The seminal work in this field was the SMS survey and no later
extensive investigation appeared in the literature during the 
last decade. The SMS
study reported the discovery of 172 candidate variables in the
central region of Carina and the search was based on  
photographic $B$ plates taken at the prime focus of the CTIO 4m
telescope. The area covered was 1 degree$^2$ centered on the 
bright field star SAO 234657, which is located very close to 
the Carina center.
Light curves and periods were obtained for 58 out of the 
73 variables identified as RR Lyrae members of the Carina
galaxy. RR Lyrae variables are considered "bona fide" tracers 
of ancient stellar populations, therefore the presence of large 
numbers of RR Lyrae provided the first evidence that an underlying 
old population was present in this galaxy. However, the mean period 
of $RRab$ stars derived by SMS is $\sim$ 0.62 days, which is 
intermediate between that of OoI ($\sim$ 0.55 days) and OoII 
($\sim$ 0.65 days) clusters. This suggests that the evolutionary 
history of the old stellar population in Carina might be different 
than that of GGCs.

Among the remaining 15 objects identified as variables by SMS, 
8 showed RR Lyrae-like light curves but magnitudes
significantly brighter than $B$=21.10 mag, the mean $B$ magnitude 
of RR Lyrae stars. Later studies (Da Costa 1988; Nemec, Nemec, 
\& Lutz 1994, and references therein) suggested that they are 
Anomalous Cepheids. These variables are
intrinsically brighter than RR Lyrae stars and follow a Period-Luminosity
(PL)  relation with the less luminous ones ($\sim$ 0.5 mag above the RR Lyrae 
level) having shorter periods ($P\sim$ 0.3 day) and the most luminous 
ones ($\sim$ 2 mag brighter than RR Lyrae stars) longer periods 
($P\sim$ 2 day). Concerning
their evolutionary status, there is a general consensus that
they are metal-poor central He-burning structures with
mass larger than 1.3$M_{\odot}$, but their origin (single young
stars or old merged binary systems) is still debated (see Bono et
al. 1997a, and references therein). In any case, they are almost
absent in GGCs (except one in NGC 5466 and two suspected candidates 
in $\omega$ Centauri), whereas they have been found in {\it all} 
dSphs that have been searched for variable stars. 
The presence of these variables might be taken as indicative of 
an underlying intermediate-age population or at least of the presence 
of a significant Blue Straggler component. The reader interested in 
a more detailed discussion is referred to paper II. 

Together with these straightforward considerations, we note that the 
radial pulsation phenomenon provides two relevant observables, period 
and luminosity amplitude, that are tightly connected with the stellar 
mass, luminosity, and effective temperature of the pulsators
(see Bono et al. 1997b; Caputo, Degl'Innocenti, \& Marconi 2002,
and references therein). These observables are affected neither
by interstellar reddening nor by distance and represent a unique
opportunity for testing the prescriptions of evolutionary and pulsation  
theories. In summary, pulsation periods and amplitudes together with  
magnitudes and colors, are the benchmarks to unambiguously 
constrain the structural parameters of variable stars, and to 
provide useful clues to understand the star-formation history 
of the host galaxy.

This paper is organized as follows: the next section addresses  
the identification of the Carina variables, while the new estimates 
of periods and mean magnitudes are presented in \S 3. Evaluations 
of the Carina distance modulus, based on the properties of RR Lyrae
stars, are discussed in \S 4. A brief summary closes the paper.

%%%%%%%%%%%%%%%%%%%%%%%%%%%%%%%%%%%%%%%%%%%%%%%%%%%%%%%%%%%%%%%%%%%%%%%%%%
\section{Identification of variables}

Observations were collected by two of us (E.B. and V.C.) over 
three consecutive nights (January 5-7, 2000) at the 2.2m ESO/MPI 
telescope (La Silla, Chile) equipped with the WFI which is a
mosaic camera with 8 EEV 2046x4098 chips. The pixel scale is
0.238 arcsec/pixel and the total field of view is $\approx$34x33
arcmin, including gaps between CCDs whose width is 23.8 and 14.3
arcsec along right ascension and declination, respectively. 
We observed with the $BV$ filters, using consecutive
exposures of $\sim$ 500 s each in order to secure a good
sampling of the light curve over the pulsation cycle. In total we
obtained 54 $BV$ pairs during the three nights. The log of the
observations is presented in Paper II together with details
concerning the seeing, the absolute, and the relative calibration,
as well as the reduction strategy.

The main aim of this investigation is to identify bright variable stars
in Carina and to estimate their pulsation parameters. To accomplish
this project we decided to perform aperture photometry and the reason
for this choice is twofold. {\em i)} The field covered by our data is 
only marginally affected by crowding, even in the innermost regions 
(see the identification maps in appendix A). 
{\em ii)} Exposure times were chosen to perform accurate photometry on 
individual frames at the luminosity typical of old HB stars, namely 
$V\approx 20.7$, $B\approx 21.0$. Data were collected with an average 
seeing better than 1 arcsec both in the B and in the V band. Therefore 
the photometry was performed using in the two bands a radius of four 
pixels. Fig. 1 shows the standard deviation as a function of the 
instrumental magnitude for the chip \# 51 of the frames \# 46388 ($b$) 
and \# 46391 ($v$) (see Table 2 in Paper II). Data plotted in this figure 
show that the photometric accuracy for an RR Lyrae variable at minimum 
light, i.e. $b\approx 15.4-15.8$ (corresponding to $B\approx 21.3-21.7$) 
and $v\approx 15.3-15.7$ (Corresponding to $V\approx 21.0-21.4$), is on 
average equal to 0.02 mag. Note that current uncertainties do not account 
for calibration errors. 

\begin{figure}
\plotone{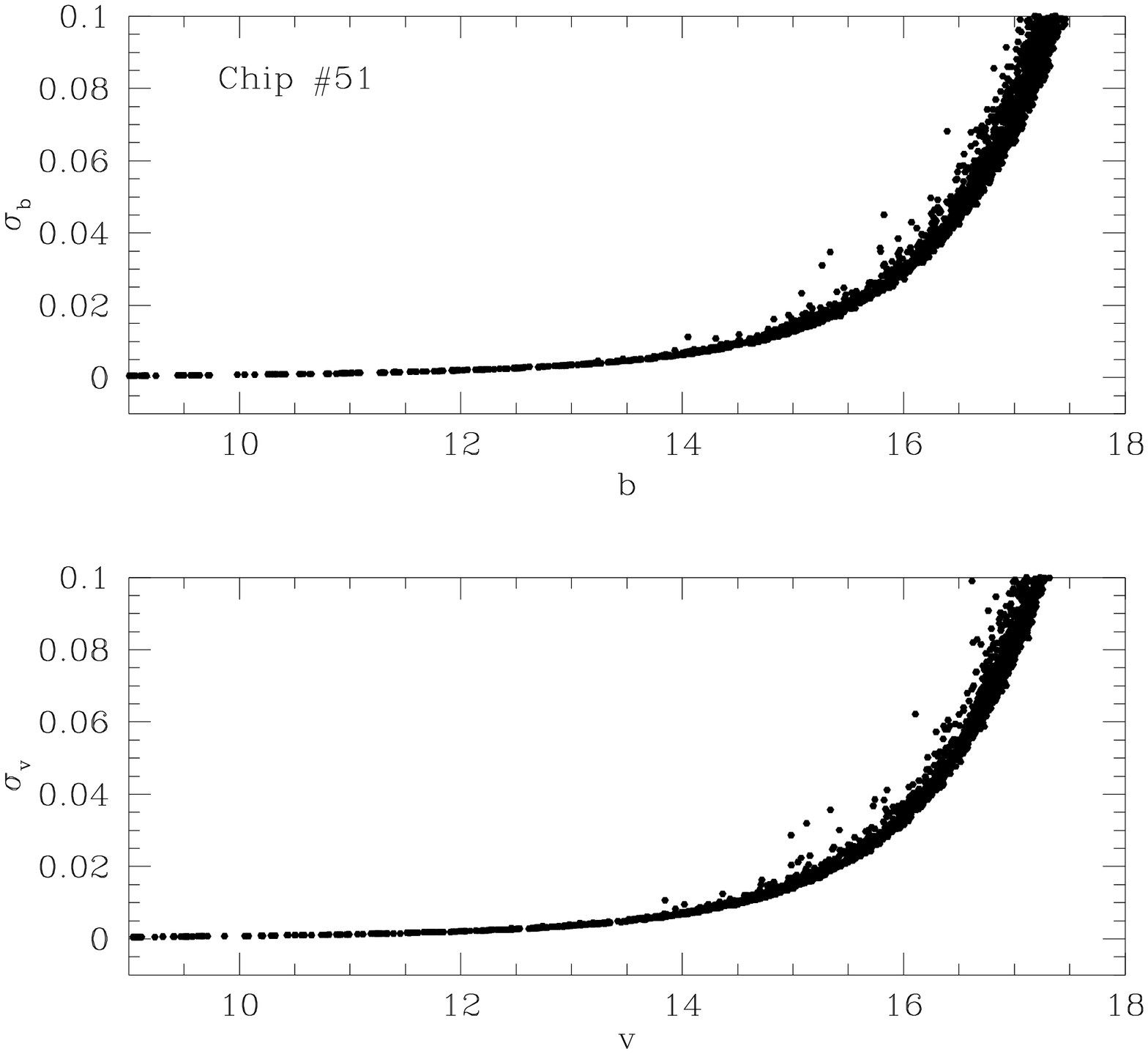}
\caption{Standard deviation as a function of the instrumental $b$ (top) 
and $v$ (bottom) magnitude. Data plotted in this figure refer to chip 
\# 51 of the frame \# 46388 ($b$) and \# 46391 ($v$), respectively. 
The exposure time of these frames is $t=500 s$, while the seeing is 
$\approx 0.8$ arcsec.\label{fig1}}
\end{figure}

\placefigure{fig1} 

The search for variable stars was carried out using the 
Welch \& Stetson (1993) algorithm. This method correlates the
magnitude variations in two or more photometric bands, and estimates 
a variability index $I$ that is very robust against spurious
effects. The higher the value of $I$, the higher is the
probability that the star is actually a "bona fide" variable star. 
We checked all the stars with $ I \geq 10$ and we found that the 
majority of variable candidates were characterized by $ I \geq 50$. 
Note that the adopted aperture photometry did not detect very faint
variable stars ($V \geq 22$ mag) and/or those characterized by very 
small luminosity amplitudes. For this reason, we plan to perform 
psf photometry on current data as well as on an additional set
of BV time series data collected by one of us (A.W.) with the 
Mosaic Imager available at the CTIO Blanco 4m telescope.

A period search was carried out on the suspected variables by means 
of a Fourier analysis. The accuracy of period evaluations  
depends on the period itself and on the time interval covered by 
observations (three nights), and it ranges from 0.004 to 0.04 
days. The phased light curves were fitted using cubic splines. 
Eventually, amplitudes and mean magnitudes (intensity-averaged 
and magnitude-averaged) were obtained by integrating the fitted 
light curves.
Table 1\footnote{The complete version of this table is only available 
in the on-line edition of the manuscript.} summarizes the results. 
The first column gives the identification 
number according to SMS, while for the new detected variables the SMS 
running number was continued, from 173 to 207. The columns 2 and 3
give $\alpha$(2000) and $\delta$(2000) coordinates, while the last two
columns list SMS and current individual notes. 
Inspection of the data in Table 1 shows that all the variables found
by SMS in the field covered by the present study, and for which
light curves and periods were obtained by those authors, have been
recovered, with the exception of 7 objects, namely V8, V9, V12, V28, 
V38, V62, and V69. The distribution of these unrecovered variables in 
the Bailey diagram (luminosity amplitude vs period) according to SMS data is 
somehow peculiar, since they present very short periods, 0.06 - 0.26 days,  
but $B$ amplitudes, 0.77 - 1.44 mag, typical of $RRab$ variables. 
According to our data they do not seem to be variable stars, and 
therefore 52 out of the 57 objects tagged as variables by SMS are 
confirmed to be variables. At the same time, only 5 out of the 81 
objects identified by SMS as suspected variables, i.e. without light 
curves and periods, have been identified as variables, namely  V87, 
V138, V141, V142, and V151. The candidate variable V161, is still 
a suspected variable.  

On the other hand, 35 new variables have been identified. This means 
that we end up with a total number of 92 detected variables. 
The individual B,V measurements for the entire sample of variables are 
listed in Table 2\footnote{The complete version of this table is only 
available in the on-line edition of the manuscript.} For each variable 
the first three columns give the Heliocentric Julian day, the B magnitude, 
and the photometric error. The columns 4,5, and 6 list the same data but for 
the V-band measurements.  
The current sample of variable stars includes 6 suspected double-mode 
pulsators (V11, V26, V84, V89, V192, V198, and V207). Note that 4 of 
them are in the SMS list, but they were classified as $RRab$ (V26, V84 V89) 
or $RRc$ (V11) variables. To firmly assess the variability of the 
7 SMS variables that did not match our variability criterion, as well 
as to properly classify the suspected multi-mode variables, new data 
that cover a longer time interval are required.  

%%%%%%%%%%%%%%%%%%%%%%%%%%%%%%%%%%%%%%%%%%%%%%%%%%%%%%%%%%%%%%%%%%%%%%%%%
\section{Periods, amplitudes,  and mean values}

Table 3 lists from left to right identification, classification, epoch,
period (days), logarithmic period, magnitude-averaged [($V$), ($B$)], 
as well as intensity-averaged [$<V>$, $<B>$] magnitudes and color [$<B>-<V>$], 
V and B amplitudes ($A_V$, $A_B$) for the 92 variables identified in the 
current photometric survey. 
Variable classification (ab=$RRab$; c=$RRc$, d=$RRd$, AC=Anomalous Cepheid) 
relies on these observables and it was a 
straightforward procedure, since different groups of radial variables 
do obey well-established relations among luminosity, temperature, 
period, and luminosity amplitude (see later).
Note that for candidate double-mode variables and for poorly-sampled 
light curves current estimates of pulsation parameters are less 
accurate. The periods of these objects are listed in italic and the 
pulsation parameters are given with two decimal figures.  
Fig. 2a-2l show the light curves of current variables. In particular  
Fig. 2a-2h display variables fainter than $<V>$=20 mag, Fig. 2i the 
suspected double-mode pulsators, Fig. 2j-2k the variables brighter than 
$<V>$=20 mag, and Fig. 2l the variables V173 and V194. The error bar 
plotted in the top left panel of Fig. 2a shows the typical B,V photometric 
uncertainty for an RR Lyrae star at minimum light. 
\placefigure{fig2a.ps-fig2l.ps}
Fig. 3 shows the location of the Carina variables in the Color-Magnitude 
diagram, derived in Paper II, according to their $<V>$ magnitudes and  
$<B>-<V>$ colors. One can easily recognize "old" HB stars extending from 
the flat RR Lyrae region ($V\sim$ 20.7 mag) to bluer colors, and the
"intermediate-age", stubby red clump stars located at $V\sim$ 20.5 mag 
and $B-V\sim$ 0.7 mag. The variables are plotted with different symbols, 
according to their position in the CMD. Specifically, the red variable 
V173 located close to the Carina RGB is marked with a cross, while the 
very bright RR Lyrae V194, probably belonging to the Galactic field, 
with a square. The remaining 90 variables are plotted using circles 
and triangles, depending on whether the apparent mean visual magnitude 
is fainter or brighter than $<V>$=20 mag.
\placefigure{fig3.ps}

Fig. 4 shows a zoom of the instability strip in the Carina CMD. 
The symbols are the same as in Fig. 3, but variables with
poorly-sampled light curves, uncertain periods, and possible 
photometric blends are displayed with open symbols. Moreover, 
for the sake of the discussion, we identify the two variables,  
V158 and V182, in the "faint" group that appear significantly 
brighter than the mean intensity-averaged magnitude, 
$<V>_m$=20.68$\pm$0.06 mag (solid line), of Carina RR Lyrae. 
A glance at the distribution of the variables plotted with
circles, and accounting for a reddening $E(B-V)=0.03\pm0.02$ mag
(see M98 and Paper II), shows that they behave as
typical globular cluster RR Lyrae stars, i.e. variables located in
the flat region of the HB with intrinsic $B-V$ colors in the range
$0.15 - 0.4$ mag (see, e.g., the extensive survey of RR Lyrae stars 
in M3 by Corwin \& Carney 2001).
\placefigure{fig4.ps}

The top panel in Fig. 5 shows the same variables plotted in the visual 
magnitude $<V>$ versus log$P$ plane (Bailey diagram). Among the variables
fainter than $<V>$=20 mag, one clearly identifies the period gap
expected between first-overtone ($RRc$) and fundamental
($RRab$) variables. On this basis, variables with log$P\le-$0.35 
and log$P\ge-$0.25 should be $c$-type (squares) and $ab$-type 
(dots) RR Lyrae, respectively. In this figure, the short-period
variables which are suspected to be double-mode pulsators are
marked with an asterisk. As already found in previous studies (M15, 
Bingham et al. 1984; IC4499, Walker \& Nemec 1996; Draco, Nemec 1985), 
they are located around the long-period edge of the $RRc$ period 
distribution.
As far as bright variables are concerned, they show periods and colors 
similar to RR Lyrae. However, their luminosity when compared with the 
luminosity of RR Lyrae becomes, on average, brighter and brighter 
toward longer periods. Bearing in mind that the luminosity increases 
toward bluer colors (see Fig. 4), one can recognize among these bright 
variables the behavior of massive ($\ge 1.3M_{\odot}$), He-burning, 
metal-poor pulsators discussed by Bono et al. (1997a). According to 
this circumstantial evidence, the variables brighter than $<V>$=20 mag 
might be tentatively classified as Anomalous Cepheids. 
Data plotted in the top panel of Fig. 5 indicate that the Anomalous 
Cepheids can be divided into two period-luminosity relations, perhaps 
signifying different pulsation modes as suggested by empirical 
(Nemec et al. 1994; Pritzl et al. 2002) and theoretical (Bono et al. 1997a) 
evidence. 
\placefigure{fig5.ps}
Note that according to this selection criterium the two variables 
V158 and V182 appear to be evolved RR Lyrae stars, since they present 
(B-V) colors quite similar to V149, the faintest AC, but their periods 
are significantly longer (see {\em infra}).    

The visual amplitude $A_V$ versus log$P$ diagram of the variables 
fainter than $<V>$=20 mag is presented in the bottom panel of Fig. 5. 
Data plotted in this panel show that the previous classification is 
confirmed. The $RRab$ candidates display a well-defined anti-correlation 
between $A_V$ and period, and indeed they attain larger amplitudes 
in the short-period range. On the other hand, the amplitude of 
$RRc$ candidates increases in the short-period ranges and decreases 
toward longer periods. It is worth noting that the almost
linear amplitude-period correlation of $ab$-type variables and the
characteristic "bell-shape" behavior of $RRc$ variables are
fully predicted by nonlinear pulsation models. According to Bono
et al. (1997b), the amplitude of fundamental pulsators increases
from the red edge (long-period) towards the blue edge (short-period) 
of the fundamental instability strip, whereas for first overtone
pulsators the amplitude attains vanishing values close to the red 
and the blue edge of its unstable region, and reach the maximum 
amplitude in the middle of the first overtone instability strip.

In the following we will try to estimate the mass range covered by
the Carina variables, based on the observed quantities inferred by
high-quality light curves and using the constraints from pulsation
theory. It is well known that the pulsation period depends on the
structural parameters (mass, luminosity, and effective temperature)
of the variable. During the last few decades the analytical relation
provided by van Albada \& Baker (1971) based on linear fundamental (F)
pulsation models

$$\log P_F=11.497+0.84\log L/L_{\odot}-0.68\log M/M_{\odot}-3.48\log
T_e\eqno(1)$$

has been the crossroad of several investigations focused on the 
evolutionary status of RR Lyrae stars.
After that pioneering approach to radial pulsation, extensive
grids of nonlinear, nonlocal, and time-dependent convective models
have been computed  (Bono et al. 1997b; Caputo et al. 2000; 
Bono et al. 2001; Bono et al. 2002; Di Criscienzo, Marconi, \& 
Caputo 2002, hereinafter DMC) for  wide ranges of metal content 
($Z=0.0001-0.02$), stellar mass ($M/M_{\odot}=0.53-0.80$) and luminosity 
($\log L/L_{\odot}=1.5-1.9$). The new theoretical scenario provides  
not only periods and modal stability across the blue edge but also 
the variation of bolometric luminosity along the pulsation cycle. 
Once the bolometric light curves are transformed into the observational 
plane using bolometric corrections and color-temperature relations  
predicted by atmosphere models (Castelli, Gratton, \& Kurucz 1997a,b), 
the nonlinear models predict observables such as amplitudes, mean
magnitudes, colors and the modal stability across the red edge. 
This means that the pulsation relation [eq. (1)] can be given in terms of 
observables, namely the mean magnitude instead of luminosity, and the 
amplitude or the mean color instead of the effective temperature.

Fig. 6 shows, according to DMC computations, the period-amplitude 
diagram of selected fundamental pulsators with the labeled mass, 
luminosity and metal content. 
Together with the well-established, and observationally confirmed, 
increase of the visual amplitude $A_V$ toward shorter 
periods\footnote{Note that, 
DMC models confirm previous findings concerning the pulsation 
properties close to the "intersection point" between first-overtone 
and fundamental blue edges (Bono, Caputo, \& Stellingwerf 1995; 
Bono et al. 1997b). Above this limit fundamental amplitudes 
present a "bell-shaped" distribution, i.e. the amplitude is vanishing 
either close to the blue and to the red edge of the instability region.}, 
one finds a dependence on both stellar luminosity (top panel) and mass 
(bottom panel). 
\placefigure{fig6.ps}
Specifically, for stellar masses ranging 
from 0.65 to 0.80$M_{\odot}$ and metal abundances from 0.0001 
to 0.001, and by adopting intensity-averaged magnitudes, DMC computations of 
fundamental models supply the following Period-Luminosity-Amplitude 
($PLA$) relation  

$$\log P_F=0.080(\pm 0.02)-0.396<M_V>-0.593\log M/M_{\odot}-0.167A_V\eqno(2)$$

It is noteworthy that the constant term and the coefficient of the $V$ 
amplitude do depend on the adopted mixing length parameter $l/H_p$=1.5. 
The effect of convection in the external layers of stellar envelopes is 
to quench pulsation driving, therefore a variation of $l/H_p$ causes 
a change in the effective temperature width of the instability region. 
The red edge of the instability strip, i.e. in the region of $ab$-type 
variables are located, is more sensitive to the adopted mixing length 
than the blue edge (Bono, Castellani, \& Marconi 2002). 
However, let us now account for a sample of $RRab$ stars located at the 
same distance, but with different absolute magnitudes and stellar masses. 
Since the period variation due to a spread in the luminosity of 
individual variables can be easily removed by adopting the "reduced" 
period, i.e. $\log P'_F$=$\log P_F$+0.396($<V>-<V>_m$) at the averaged 
observed magnitude $<V>_m$, the "reduced" Bailey diagram 
should give information on the spread in mass of individual variables. 
For example, by adopting a mass range of $M=0.65-0.80 M/M_{\odot}$, 
the expected dispersion in the "reduced" period, at constant $A_V$, is
$\Delta$log$P'_F\sim$0.09. 

Fig. 7 shows the $A_V$ versus log$P'_F$ diagram for the Carina 
$RRab$ stars, together with similar diagrams for variables in M3 
($<V>_m$=15.62 mag, Corwin \& Carney 2001) and M15
($<V>_m$=15.82 mag, Bingham et al. 1984). We selected these two 
clusters because they host sizable samples of RR Lyrae stars (207 in M3,  
88 in M15) for which accurate pulsation parameters are available, 
and they are the typical OoI (M3) and OoII (M15) Galactic prototypes.
Moreover, they bracket the Carina mean metallicity, i.e. [Fe/H]=-1.7 
(see Paper II), while the mean metallicity for M3 and M15 are 
[Fe/H]=-1.3, and -2.15 (Harris 1996), respectively. 
The dashed lines plotted in Fig. 7 show a period dispersion at constant 
amplitude of $\Delta$log$P'_F$=0.09, as estimated using eq. (2) and by 
assuming a mass range of $M=0.65-0.80 M_{\odot}$. It is noteworthy that 
all the variables in the three stellar systems fulfill this assumption, 
with the exception of  V158 and V182 in Carina which appear to have a 
significantly shorter reduced period.
According to eq. (2), we estimate that their mass is larger by
$\sim$50\% than the bulk of Carina $RRab$ variables.
\placefigure{fig7.ps}

A further observable which is related to the effective temperature
is obviously the $B-V$ color. Note that
the observed color is not the static color, i.e the color of the
variable if it were a static star. Several empirical investigations  
and the data listed in Table 3 show that the magnitude-averaged colors
($B-V$) are redder than the intensity-averaged colors ($<B>-<V>$). 
This difference for $RRab$ stars increases when moving from low 
(long period) to high-amplitude (short period) pulsators.
Nonlinear pulsation models (Bono, Caputo \& Stellingwerf 1995;
DMC) confirm such a behavior, and also provide the
amplitude-correction to observed mean colors to estimate  
the intrinsic static $(B-V)_s$ color. Given the typical
distribution of RR Lyrae variables in the period-amplitude diagram, 
one clearly understands that this correction for $RRc$ is quite small,
whereas for the shorter period $RRab$ variables it becomes a sizable 
correction. In particular, for $(B-V)_s$=0.27 mag and $A_V\sim$ 1.5 mag, 
one finds $<B>-<V>\approx$ 0.24 mag and ($B-V$)$\approx$ 0.31 mag. 
Note also that the mean $V$ magnitudes are always fainter than the 
static value $V_s$, again with the
difference increasing with $A_V$. With $A_V\sim$ 1.5, one has
$V_s-(V)\sim -$0.14 mag and $V_s-<V>\sim -$0.04 mag. As a  
consequence, in the region of the instability strip where the 
transition between the two pulsation modes takes place, $RRc$ 
and $RRab$ variables with similar parameters (mass,
luminosity, and effective temperature) may have significatively 
different mean magnitudes and colors. Once the photometric 
database of this project will be completed we plan to derive 
static magnitudes and colors for all the Carina variables and 
to perform a detailed analysis. 
At present, since the nonlinear pulsation models provide directly 
information on the mean magnitudes, from DMC computations with 
0.0001$<Z<$0.001 we derive the following Period-Luminosity-Color 
($PLC$) relation 

$$\log P_F=-0.553(\pm 0.014)-0.617\log M/M_{\odot}-0.347<M_V>
+(1.007-0.070\log Z)[<B>-
<V>]\eqno(3)$$

This equation also accounts for first overtone pulsators, provided that 
the period has been "fundamentalized", i.e. log$P_F$=log$P_{FO}$+0.13. 
According to this equation, a homogeneously reddened sample of RR
Lyrae stars located at the same distance and with roughly constant
mass and metallicity, but different luminosity and color, should
have a well-established locus in the log$P'_F$
(=log$P_F$+0.347($<V>-<V>_m$) versus $<B>-<V>$ plane. On the contrary,
if the stellar mass of RR lyrae stars is significantly different, then
the more massive variables should present, at fixed color, shorter 
reduced periods, and vice versa.

Fig. 8 shows the reduced periods versus dereddened color for RR Lyrae
stars in Carina, M3 and M15. Also in this case, the dashed lines
depict the predicted period dispersion at constant color
($\Delta$log$P'_F\sim$0.09) by assuming the mass range of 
$M=0.65-0.80 M_{\odot}$. Data plotted in this figure show that all the
variables in M3 and M15 are quite consistent with the adopted mass
range, whereas the RR Lyrae stars in Carina seem to suggest a
slightly larger range. This hypothesis will be checked using static
values. In any case, present data plotted in Fig. 8 confirm that
V158 and V182 are significantly more massive (by $\sim$ 50\%) than
the bulk of Carina RR Lyrae stars.
\placefigure{fig8.ps}

We can now take into account the variables brighter than $<V>$=20 mag.
Fig. 9 shows the comparison between variables with good-quality 
light curves and the mean locus of RR Lyrae stars plotted 
in Fig. 8 (dashed line). Data plotted in this figure disclose 
that their reduced periods are shorter than those of V158 and V182, 
thus suggesting even larger masses. 
As a whole, the Carina bright variables attain reduced periods that 
are systematically shorter than the predicted values  
$\Delta \log P'_F=-$0.11 and $-0.25$ (solid lines) provided by eq. (3) 
with mass values that are 1.5 and 2.5 times larger than the bulk of 
fundamental variables fainter than $<V>$=20 mag. Note that similar 
mass estimates for ACs have been derived in Paper II on the basis 
of the comparison between Zero Age He-burning structures and empirical 
data. However, since previous studies (Nemec, Nemec, \& Lutz 1994; Bono
et al. 1997a) have suggested the presence in dSphs of both
fundamental and first overtone Anomalous Cepheids, a more detailed  
analysis is required to constrain the pulsation mode, and in turn 
to estimate the individual mass of these variables.
\placefigure{fig9.ps}

Finally, let us compare the pulsation properties of "bona fide" 
Carina RR Lyrae stars (with the exception of V158 and V182) with 
variables observed in other dSph galaxies and globular clusters in 
the Milky Way. 
According to the data listed in Table 3, we derived $<\log Pab>=-0.200$ 
($<Pab>=0.631$ days) and $N_{cd}/(N_{ab}+N_{cd})$=0.29, 
where $N_{cd}$ includes the suspected $RRd$ variables. 
where $N_{ab}$ is the number of $RRab$ variables as well as $N_{cd}$ 
the number of $RRc$ and suspected $RRd$ variables. 
Data plotted in Fig. 10 disclose 
that, the mean period of $RRab$ variables in Carina (filled triangle) 
appear quite similar to the values observed in RR Lyrae-rich OoII clusters 
(filled circles), whereas the fraction of $RRc$ is significantly smaller 
when compared with those clusters and close to the values observed in 
OoI systems (open circles). Current data for RR Lyrae in dSphs (open 
triangles, see also Table 4), confirm that these stellar systems can 
be hardly classified into the "classical" Oosterhoff groups. 
\placefigure{fig10.ps}
In fact, Ursa Minor presents a mean fundamental period 
($\log <P_{ab}>=-0.197$) and a fraction of $RRc$ stars $\approx 0.4$,  
that are typical of Oo type II clusters, while Sagittarius has  
$\log < P_{ab}>=-0.241$ and a fraction of $RRc$ stars 
$\approx 0.2$, that are typical of Oo type I clusters.  
Moreover, Fig. 11 shows the comparison between the period distribution 
of $RRab$ stars in Carina with the cumulative period distribution 
of three OoII clusters with sizable samples of RR Lyrae, namely M15
(Cox, Hodson, \& Clancy 1983; Silbermann, \& Smith 1995), M53 (Kopacki 2000), 
and M68 (Walker 1994). For these clusters current data suggest $N_{ab}$=49, 
$\log P_{ab}>=-0.187$ ($<P_{ab}>=0.653$ days), and 
$N_{cd}/(N_{ab}+N_{cd})\sim0.50$.
Interestingly enough, the Carina variables show a gaussian distribution 
with a peak near $\log P\sim-0.20$, whereas the cluster variables show 
a bimodal distribution with a long-period peak at $\sim-0.14$. 
Whether this difference might be due to the fact that Carina shows a 
redder HB morphology than OoII clusters it is not clear. However, this 
might mean that Carina lacks of significant numbers of bright variables 
evolving redward from the blue side of the RR Lyrae instability strip 
(Caputo, Tornamb\'e \& Castellani 1989). Note that this working hypothesis 
could explain the low fraction of $RRc$ variables detected in Carina. 
\placefigure{fig11.ps}

%%%%%%%%%%%%%%%%%%%%%%%%%%%%%%%%%%%%%%%%%%%%%%%%%%%%%%%%%%%%%%%%%%%%%%%%%%%%%
\section{RR Lyrae stars as distance indicators}

RR Lyrae stars have been one of the most popular distance indicator 
for metal-poor stellar systems, since their intrinsic luminosity 
is expected to mainly depend on the metal content [Fe/H]. As a 
consequence, in the last few decades many investigations have 
been devoted to the slope and the zero-point of the $M_V$(RR) 
versus [Fe/H] relation. Even though paramount theoretical and 
observational efforts have been devoted to this long-standing 
problem a broad consensus on "long" and "short" distance scales 
has not been reached yet.  

Among the different approaches discussed in the recent literature, 
we focus our attention on those based on the analysis of pulsation 
properties.  These methods are attractive, since they rely on 
observables (period and luminosity amplitude) that are not affected 
by uncertainties on both distance modulus and reddening.
In this context, Caputo (1997) and Caputo et al. (2000) developed 
a method based on the predicted PL relation for 
pulsators located along the first overtone blue edge (FOBE), that 
seems quite robust for clusters with significant numbers of $RRc$ 
variables. According to this procedure, for each assumption concerning 
the globular cluster distance modulus, one obtains the distribution 
of the cluster RR Lyrae stars in the $M_V$-log$P$ plane. A determination 
of the actual distance modulus is derived by matching the observed 
distribution of $RRc$ variables with the predicted relation for the  
FOBE (Caputo et al. 2000), i.e.  

$$M_V(FOBE)=-0.685(\pm 0.027)-2.255\log P(FOBE)-1.259\log M/M_{\odot}+0.058\log Z\eqno(4)$$

in such a way that no $RRc$ variables are located in the hotter stable 
region. 

By adopting for Carina a metallicity of $[Fe/H]=-1.7$ (i.e. 
$\log Z=-3.4$, see Paper II for details), we can estimate on the basis  
of evolutionary HB models that the mass of $RRc$ variables is 
$M=0.7 M_{\odot}$, with an uncertainty of the order of 4\% 
(see Bono et al. 2002). On this basis, the FOBE procedure 
(see Fig. 12) yields an apparent distance modulus for Carina 
of $DM_V$=20.19$\pm$0.04 mag.
\placefigure{fig12.ps}

The distance to Carina can also be estimated using eq. (2) 
and by adopting $M=0.75 M_{\odot}$ for $RRab$ variables 
(Bono et al. 2002). This choice is based on the evolutionary evidence 
that the mass of HB models increases towards the red, i.e. when 
moving from $RRc$ to $RRab$ variables. As a result, we derive 
$DM_V$=20.09$\pm$0.10 mag.

Finally, eq. (3) can also be adopted to estimate the distance to Carina. 
In particular, by assuming $M=0.7 M_{\odot}$ for $RRc$ stars and 
$0.75 M_{\odot}$ for $RRab$, we obtain $DM_V$=20.19$\pm$0.12 mag.
Bearing in mind that systematic effects due to the adopted mixing-length 
parameter may affect the predicted $PLA$ relation, our final estimate 
of the Carina distance modulus according to pulsational methods is 
$DM_V$=20.19$\pm$0.12 mag. This yields $<M_V(RR)>$=0.49$\pm$0.12 mag 
that is slightly fainter than the value 0.42$\pm$0.08 mag inferred from 
the $<M_V(RR)>$-[Fe/H] relation given by Bono et al. (2002) for metal-poor 
RR Lyrae stars. On the other hand, the Bono et al. relation for [Fe/H]=-1.5
gives $<M_V(RR)>=0.50\pm0.09$ mag, that is slightly brighter than the 
estimate recently provided by Cacciari (2002), $<M_V(RR)>=0.57\pm0.04$ 
at [Fe/H]=-1.5, on the basis of several theoretical and empirical 
calibrations.

%%%%%%%%%%%%%%%%%%%%%%%%%%%%%%%%%%%%%%%%%%%%%%%%%%%%%%%%%%%%%%%%%%%%%%%%%%%
%				Summary 
%%%%%%%%%%%%%%%%%%%%%%%%%%%%%%%%%%%%%%%%%%%%%%%%%%%%%%%%%%%%%%%%%%%%%%%%%%%
\section{Summary and conclusions}

We collected BV time series data in a region of $34\times33$ arcmin 
centered on the Carina dSph with the WFI available at the 2.2m ESO/MPI  
telescope. In this investigation we present some preliminary results 
concerning bright variables located inside the Cepheid instability strip, 
namely RR Lyrae and Anomalous Cepheids. The observing strategy and the 
data reduction approach allowed us to identify 92 variables. Among 
them 75 are RR Lyrae and 15 are "bona fide" Anomalous Cepheids. We also 
found a field RR Lyrae and a variable located close to the Carina RGB. 
Current light curves when compared with those derived by SMS on the basis 
of photographic plates present, as expected, smooth variations over the 
entire pulsation cycle. Among the 75 RR Lyrae stars, 26 are new 
identifications, while 6 are new candidate double-mode RR Lyrae, while 
among the 15 Anomalous Cepheids, 7 are new identifications. 
Moreover, we recovered 57 out of the 140 variable sources (candidates plus 
"bona fide" variables) detected by SMS in the field covered by our data.  
However, the aperture photometry we adopted in this investigation does 
not allow us to assess on a firm basis the complete census of variable 
sources in this field. Moreover, it is worth mentioning that the current 
sample of bright variables is also biased toward shorter periods, since 
current data cover a time interval of 3 days. To overcome these problems 
we plan to perform PSF photometry and to merge the photometric databases 
collected with the 2.2m and with the CTIO 4m telescopes.  
 
Current data concerning the pulsation properties of RR Lyrae 
variables disclose that they are quite similar to RR Lyrae in 
GGCs. Interestingly enough, we found 
that two objects, V158 and V182, 
out of the 75 variables fainter than 21.1 mag might be more massive 
than typical RR Lyrae stars. Their reduced periods are systematically 
shorter than the bulk of the RR Lyrae stars. 
Straightforward physical arguments based on the pulsation 
properties of bright variables, i.e. $V\le 20$ mag strongly support 
the evidence that they are Anomalous Cepheids, i.e. He burning stars 
with stellar masses 2 times more massive than RR Lyrae stars.  
Further speculations concerning the pulsation mode, fundamental or 
first overtone, as well as new insights on their evolutionary 
history, single or binary (Bono et al. 1997a) require an accurate 
coverage of the light curve and detailed star count estimates of 
the intermediate-age main sequence stars. 

We adopted three different methods to evaluate the Carina distance 
modulus. The FOBE method suggested by Caputo et al. (2000) provides,
assuming a mean metal content of [Fe/H]=-1.7, an apparent distance 
modulus of $20.19\pm0.04$ mag, the Period-Luminosity-Amplitude relation  
gives, assuming mean masses of $M=0.7M_\odot$ 
for $RRc$ and $M=0.75M_\odot$ for $RRab$ variables, $20.09\pm0.10$
mag, while the period-luminosity-color relation supplies $20.19\pm0.12$ mag.
As a whole, our distance modulus evaluation based on RR Lyrae stars is  
$20.19\pm0.12$ mag, and assuming a reddening correction of 
$E(B-V)=0.03\pm0.02$ (see paper II) we derive true distance modulus of 
$\mu_0=20.10\pm0.12$. Distance estimates available in the literature 
present a large spread and depend on the adopted method. They range   
from $\mu_0=19.87$ (Mighell 1997) to $\mu_0=20.19\pm0.13$ (Dolphin 2002) 
according to the CMD fitting. On the other hand, Smecker-Hane et al. (1994)
found $\mu_0=20.05\pm0.06$ and $\mu_0=20.12\pm0.08$ using the Tip of the 
RGB and the luminosity of HB stars, respectively. A similar 
distance ($\mu_0=20.06\pm0.12$) was found by Mateo, Hurley-Keller, 
\& Nemec (1998) using the Period-Luminosity relation for dwarf Cepheids 
derived by McNamara (1995). The quoted authors typically assumed a 
mean metallicity of $[Fe/H]=-2.2\pm0.2$, and a mean reddening of 
$E(B-V)\approx0.03\pm0.02$. More recently, Girardi \& Salaris (2001) 
provided an independent distance evaluation using a new calibration of 
the I-band absolute magnitude of the red giant clump and the apparent 
I magnitude of the Carina clump derived by Udalski (1998) and for 
[Fe/H]=-1.7, $E(B-V)\approx0.06$ they found $\mu_0=19.96\pm0.06$ mag. 
More accurate constraints on the Carina true distance modulus 
will be provided as soon as the search for variable stars will 
extend to fainter variables such as $\delta$ Scuti and 
Oscillating Blue Stragglers. The dSph galaxies 
that host stellar population with a wide spread in stellar ages 
are a fundamental laboratory to constrain the intrinsic accuracy  
of different groups of standard candles (Bono et al. 2002). 

The data for Carina together with data for RR Lyrae variables in dSphs 
available in the literature support the conclusion that these stellar 
systems cannot be easily classified into either of the classical 
Oosterhoff groups. In the case of Carina, the mean period of the $RRab$ 
variables is quite similar to the mean period of $RRab$ stars in Oosterhoff 
type II clusters. On the other hand, the ratio between $RRc$ and total 
number of RR Lyrae stars is typical of that seen in Oosterhoff type I 
globular clusters. Carina shares these characteristics with some not all 
of the other well-studied dSphs.

\acknowledgments  
We wish to thank the referee, Dr. J. Nemec, for several pertinent 
suggestions that improved the content and the readibility of the paper.  
This work was partially supported by MURST/COFIN~2000 under the 
project: "Stellar Observables of Cosmological Relevance". HAS thanks 
the US National Science Foundation for support under the grant AST99-86943.  

%%%%%%%%%%%%%%%%%%%%%%%%%%%%%%%%%%%%%%%%%%%%%%%%%%%%%%%%%%%%%%%%%%%%%%%
%                               Appendix
%%%%%%%%%%%%%%%%%%%%%%%%%%%%%%%%%%%%%%%%%%%%%%%%%%%%%%%%%%%%%%%%%%%%%%%
\clearpage \appendix
\section{Comments on individual variables} 

The identification maps for the entire sample of variable stars are only 
available in the on-line edition of the manuscript (see Figures 13 to 20). 
\placefigure{fig13.ps-fig20.ps} 

V10- It is located close to the spike of a bright star. 

V11- It is located close to a bright star. The light curve is noisy, but 
it is not clear whether it is a double-mode pulsator. 

V14- The light curve is mildly noisy. Possible blends with nearby faint stars. 

V26- Possible double-mode pulsator. 

V27- Poor-sampled light curve.  

V68- The light curve is mildly noisy, possible blends with nearby faint stars.  

V74- The B-band light curve is noisy. It is located close to the edge of 
the chip \# 51.

V84- The light curve is noisy. It is located very close to a bright star.  

V89- Possible double-mode pulsator.

V115- Poor-sampled light curve. 

V122- Poor-sampled light curve. Possible blend with nearby faint stars. 

V129- Poor-sampled light curve. It is located close to the edge of the 
chip \# 55.  

V142- It is located close to a bright star. Possible blend.  

V149- Poor-sampled light curve. 

V153- The light curve is noisy. Located close to a bright star. 

V164- Located close to the edge of the chip \# 53. Possible blend with 
a field galaxy.

V176- Poor-sampled light curve. It is located close to a bad column 
and are available 33 out of 54 measurements. 

V177- Poor-sampled light curve. The star is located close to the edge of 
the chip \# 50. 

V178- Poor-sampled light curve. Located close to the edge of the chip \# 51. 

V180- Poor-sampled light curve. 

V181- The light curve is mildly noisy. Possible blend with a faint star. 

V182- Possible blend with a faint star. 

V187- Poor-sampled light curve.  

V190- Poor-sampled light curve.  

V192- Possible double-mode pulsator or a blend with nearby faint stars. 

V198- Possible double-mode pulsator.  

V200- Possible blend with a faint star. 

V203- Poor-sampled light curve. Possible blend with a nearby faint star. 

V207- Possible double-mode pulsator.

%%%%%%%%%%%%%%%%%%%%%%%%%%%%%%%%%%%%%%%%%%%%%%%%%%%%%%%%%%%%%%%%%%%%%%%%%%%%%%%
%					REFERENCES			     
%%%%%%%%%%%%%%%%%%%%%%%%%%%%%%%%%%%%%%%%%%%%%%%%%%%%%%%%%%%%%%%%%%%%%%%%%%%%%%%
\pagebreak

%%%%%%%%%%%%%%%%%%%%%%%%%%%%%%%%%%%%%%%%%%%%%%%%%%%%%%%%%%%%%%%%%%%%%%%%%%%%%%
%					TABLES			     %
%%%%%%%%%%%%%%%%%%%%%%%%%%%%%%%%%%%%%%%%%%%%%%%%%%%%%%%%%%%%%%%%%%%%%%%%%%%%%%
\footnotesize{
\include{dallora.table1}
\include{dallora.table3}
\include{dallora.table4}
}

%%%%%%%%%%%%%%%%%%%%%%%%%%%%%%%%%%%%%%%%%%%%%%%%%%%%%%%%%%%%%%%%%%%%%%%%%%%%%
%	FIGURE CAPTIONS			     %
%%%%%%%%%%%%%%%%%%%%%%%%%%%%%%%%%%%%%%%%%%%%%%%%%%%%%%%%%%%%%%%%%%%%%%%%%%%%%

%Fig 1
\clearpage

%Fig 2a 
%\clearpage
\begin{figure}
\plotone{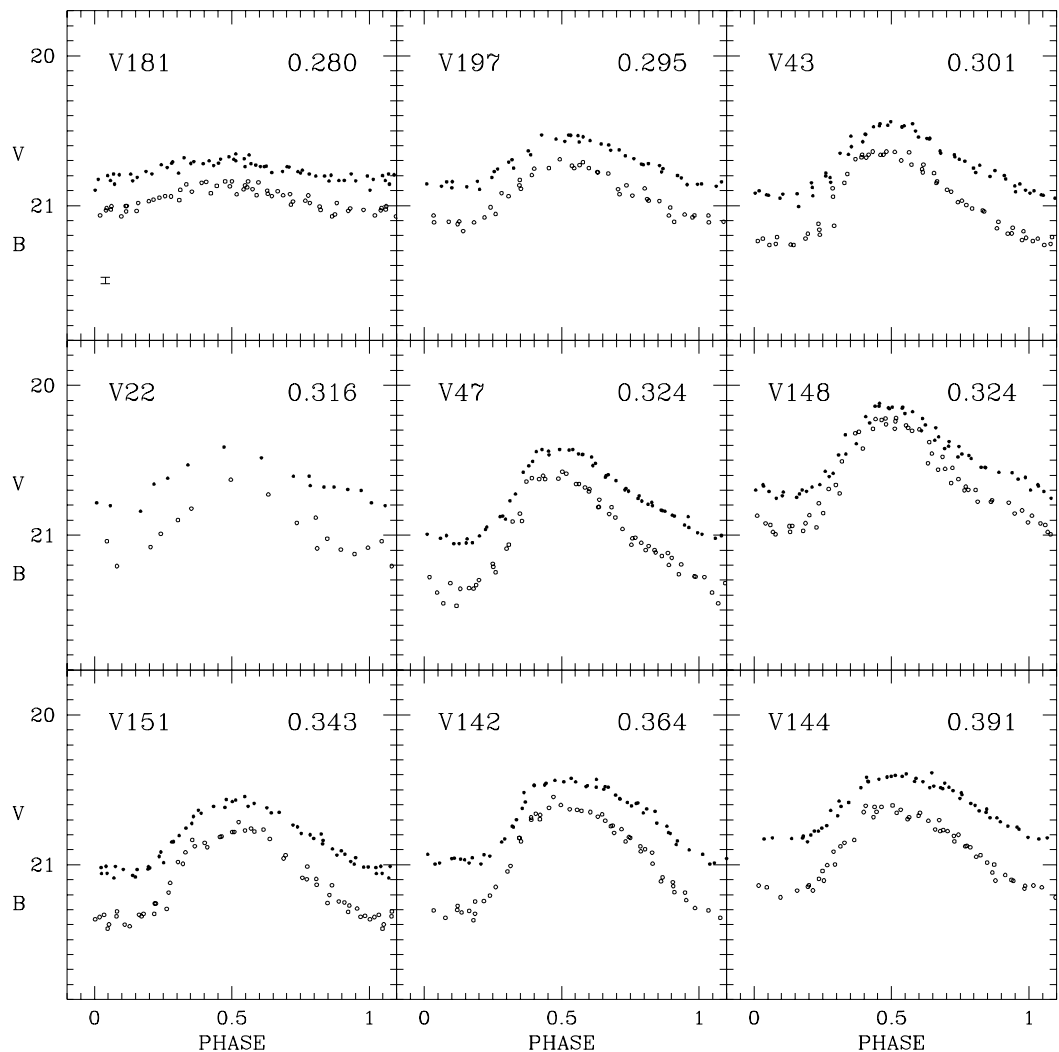}
\newcounter{mio}
\setcounter{mio}{1}
\renewcommand{\thefigure}{\arabic{figure}-\alph{mio}}
\caption{Light curves of fundamental and first overtone RR Lyrae detected 
in the central region of the Carina dSph. Filled and open squares display 
V and B-band light curves, respectively. In each panel are listed the 
identification and the period (d). The error bar shows the typical 
photometric uncertainty for an RR Lyrae at minimum light.\label{fig2a}}
\end{figure}

%Fig 2b 
%\clearpage
\begin{figure}
%\plotone{dallora.fig2b.ps}
\setcounter{mio}{2}
\setcounter{figure}{1}
\renewcommand{\thefigure}{\arabic{figure}-\alph{mio}}
\caption{Continued.\label{fig2b}}
\end{figure}

%Fig 2c 
%\clearpage
\begin{figure}
%\plotone{dallora.fig2c.ps}
\setcounter{mio}{3}
\setcounter{figure}{1}
\renewcommand{\thefigure}{\arabic{figure}-\alph{mio}}
\caption{Continued.\label{fig2c}}
\end{figure}

%Fig 2d 
%\clearpage
\begin{figure}
%\plotone{dallora.fig2d.ps}
\setcounter{mio}{4}
\setcounter{figure}{1}
\renewcommand{\thefigure}{\arabic{figure}-\alph{mio}}
\caption{Continued.\label{fig2d}}
\end{figure}

%Fig 2e 
%\clearpage
\begin{figure}
%\plotone{dallora.fig2e.ps}
\setcounter{mio}{5}
\setcounter{figure}{1}
\renewcommand{\thefigure}{\arabic{figure}-\alph{mio}}
\caption{Continued.\label{fig2e}}
\end{figure}

%Fig 2f
%\clearpage
\begin{figure}
%\plotone{dallora.fig2f.ps}
\setcounter{mio}{6}
\setcounter{figure}{1}
\renewcommand{\thefigure}{\arabic{figure}-\alph{mio}}
\caption{Continued.\label{fig2f}}
\end{figure}

%Fig 2g  
%\clearpage
\begin{figure}
%\plotone{dallora.fig2g.ps}
\setcounter{mio}{7}
\setcounter{figure}{1}
\renewcommand{\thefigure}{\arabic{figure}-\alph{mio}}
\caption{Continued.\label{fig2g}}
\end{figure}

%Fig 2h  
%\clearpage
\begin{figure}
%\plotone{dallora.fig2h.ps}
\setcounter{mio}{8}
\setcounter{figure}{1}
\renewcommand{\thefigure}{\arabic{figure}-\alph{mio}}
\caption{Continued.\label{fig2h}}
\end{figure}

%Fig 2i
%\clearpage
\begin{figure}
%\plotone{dallora.fig2i.ps}
\setcounter{mio}{9}
\setcounter{figure}{1}
\renewcommand{\thefigure}{\arabic{figure}-\alph{mio}}
\caption{Same as Fig. 2a, but the light curves refer to the suspected 
double-mode variables.\label{fig2i}}
\end{figure}

%Fig 2j
%\clearpage
%\pagebreak
\begin{figure}
%\plotone{dallora.fig2j.ps}
\setcounter{mio}{10}
\setcounter{figure}{1}
\renewcommand{\thefigure}{\arabic{figure}-\alph{mio}}
\caption{Same as Fig. 2a, but the light curves refer to Anomalous Cepheids.
\label{fig2j}}
\end{figure}

%Fig 2k
%\clearpage
\begin{figure}
%\plotone{dallora.fig2k.ps}
\setcounter{mio}{11}
\setcounter{figure}{1}
\renewcommand{\thefigure}{\arabic{figure}-\alph{mio}}
\caption{Continued.\label{fig2k}}
\end{figure}

%Fig 2l
%\clearpage
\begin{figure}
%\plotone{dallora.fig2l.ps}
\setcounter{mio}{12}
\setcounter{figure}{1}
\renewcommand{\thefigure}{\arabic{figure}-\alph{mio}}
\caption{Light curve of the variable V173, located close to the Carina RGB,
and of the variable V194, a possible Galactic field RR Lyrae.\label{fig2l}}
\end{figure}

%Fig 3
\clearpage
\begin{figure}
\plotone{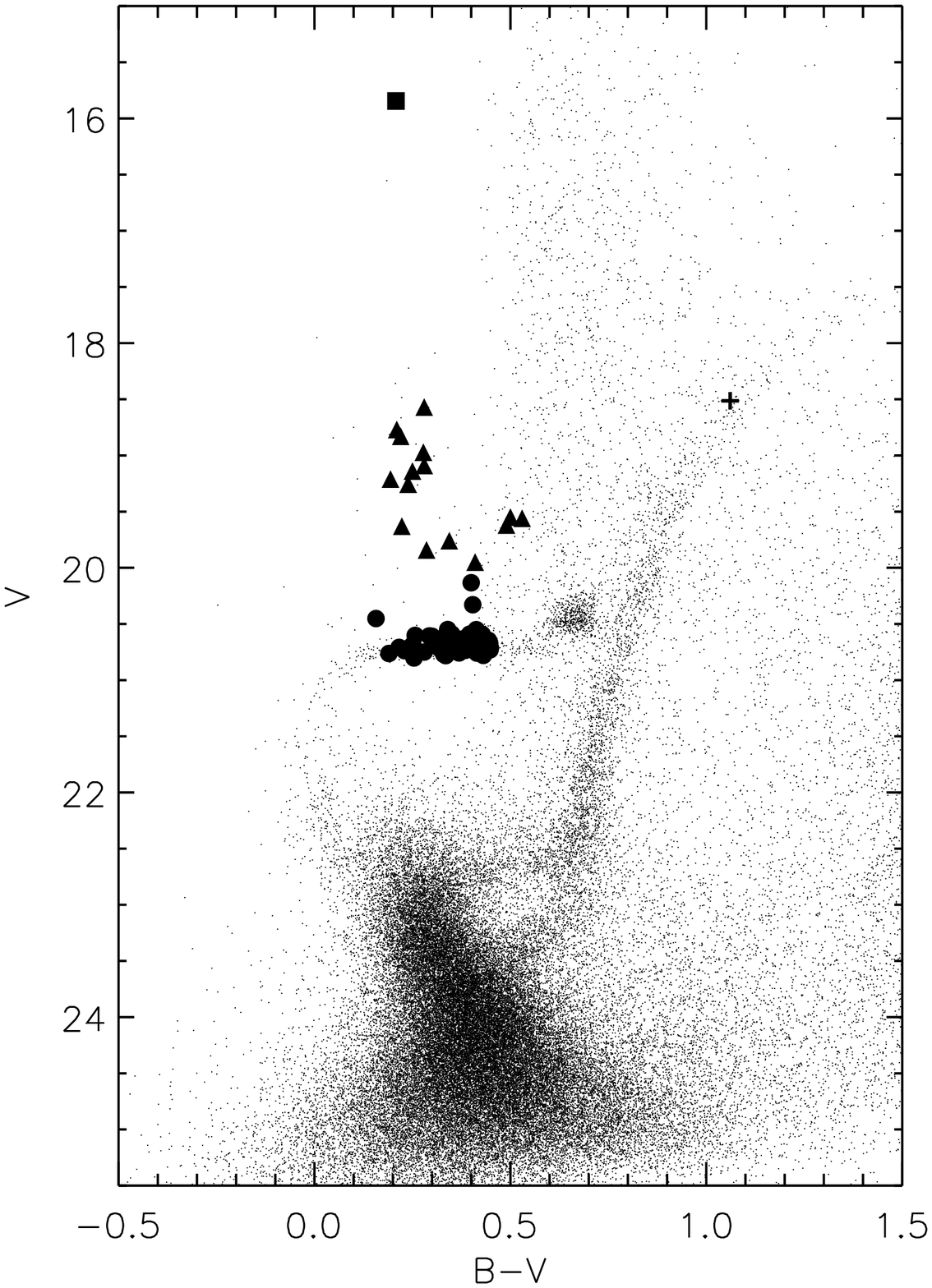}
\bigskip
\bigskip
\bigskip
\caption{Location of the 92 identified variables in the V,B-V 
color-magnitude diagram. Circles and triangles mark variables fainter 
and brighter than $<V> = 20$ mag respectively. Variable V173 is shown 
with a cross, and variable V194 with a square (see text for 
details).}
\end{figure}

%Fig 4
\clearpage
\begin{figure}
\plotone{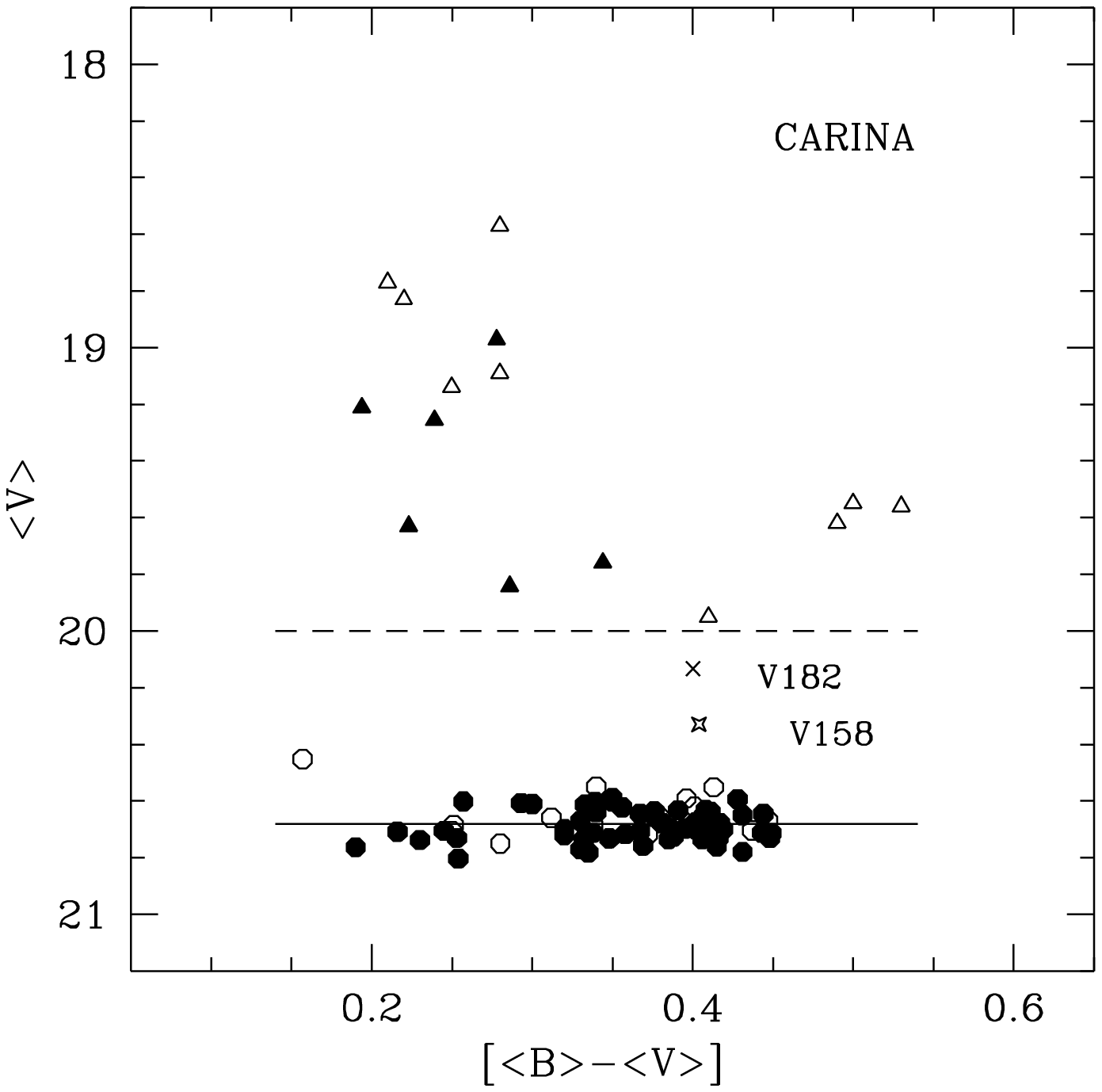}
\caption{Close-up of the instability strip in the CM diagram. 
The symbols are the same as in Fig. 3, but variables with poorly-sampled 
light curves, uncertain periods, or possible blends are plotted with 
open symbols. The dotted line marks the edge between faint and bright 
variables. The variables V158 and V182 that appear significantly brighter 
than the mean intensity-averaged magnitude of the bulk of RR Lyrae stars 
($<V>_m=20.68$ mag, solid line) are displayed with special symbols.}
\end{figure}

%Fig 5
\clearpage
\begin{figure}
\plotone{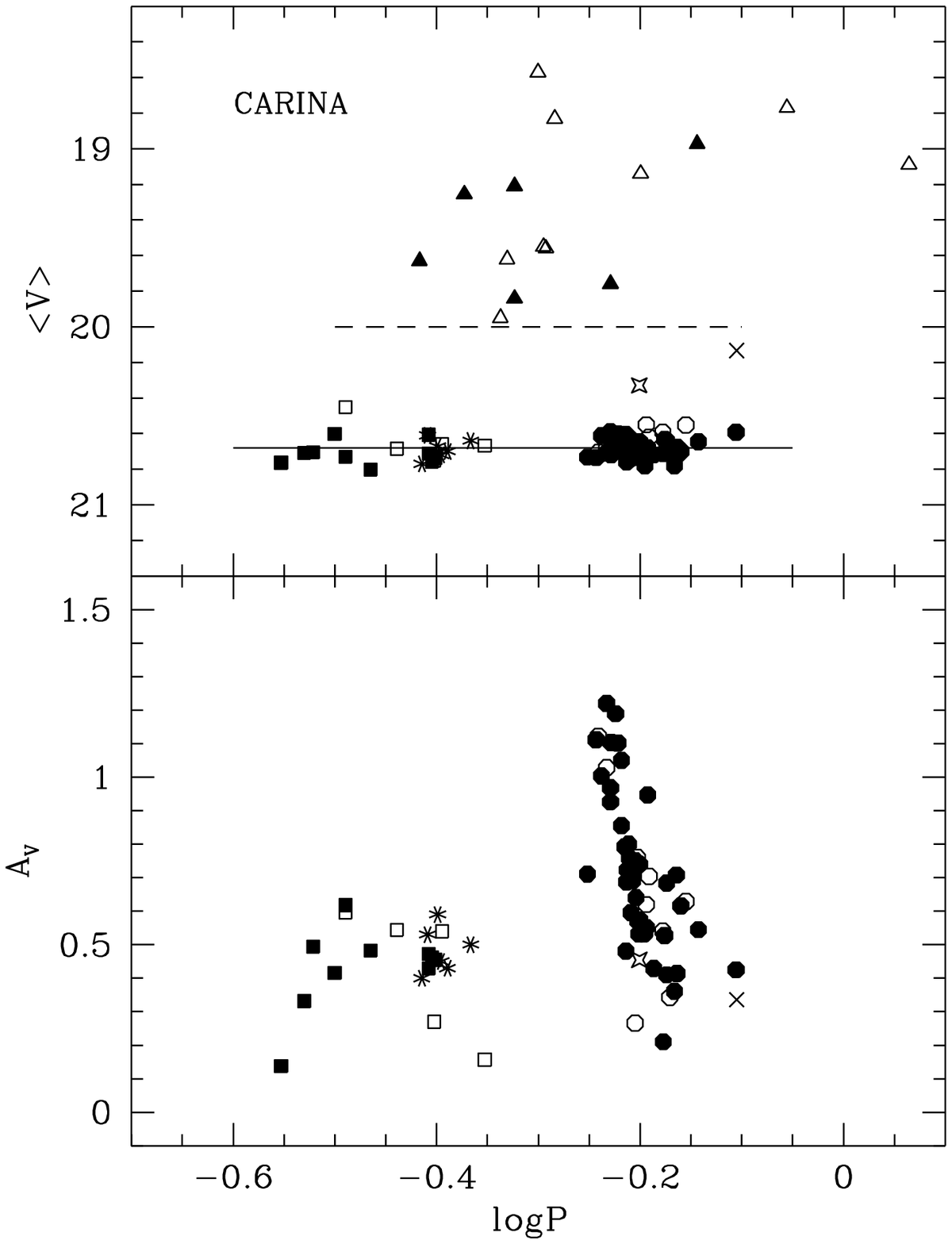}
\caption{Top: Carina variables in the visual magnitude
$<V>$ vs log$P$ plane. The gap between fundamental (circles) and first 
overtone (squares) variables  is clearly visible. The asterisks mark 
the suspected double-mode pulsators. The open symbols show variables 
with uncertain properties.
Bottom: Carina variables in the Bailey diagram, V amplitude vs log$P$. 
The symbols are the same as in the top panel.}  
\end{figure}

%Fig 6
\clearpage
\begin{figure}
\plotone{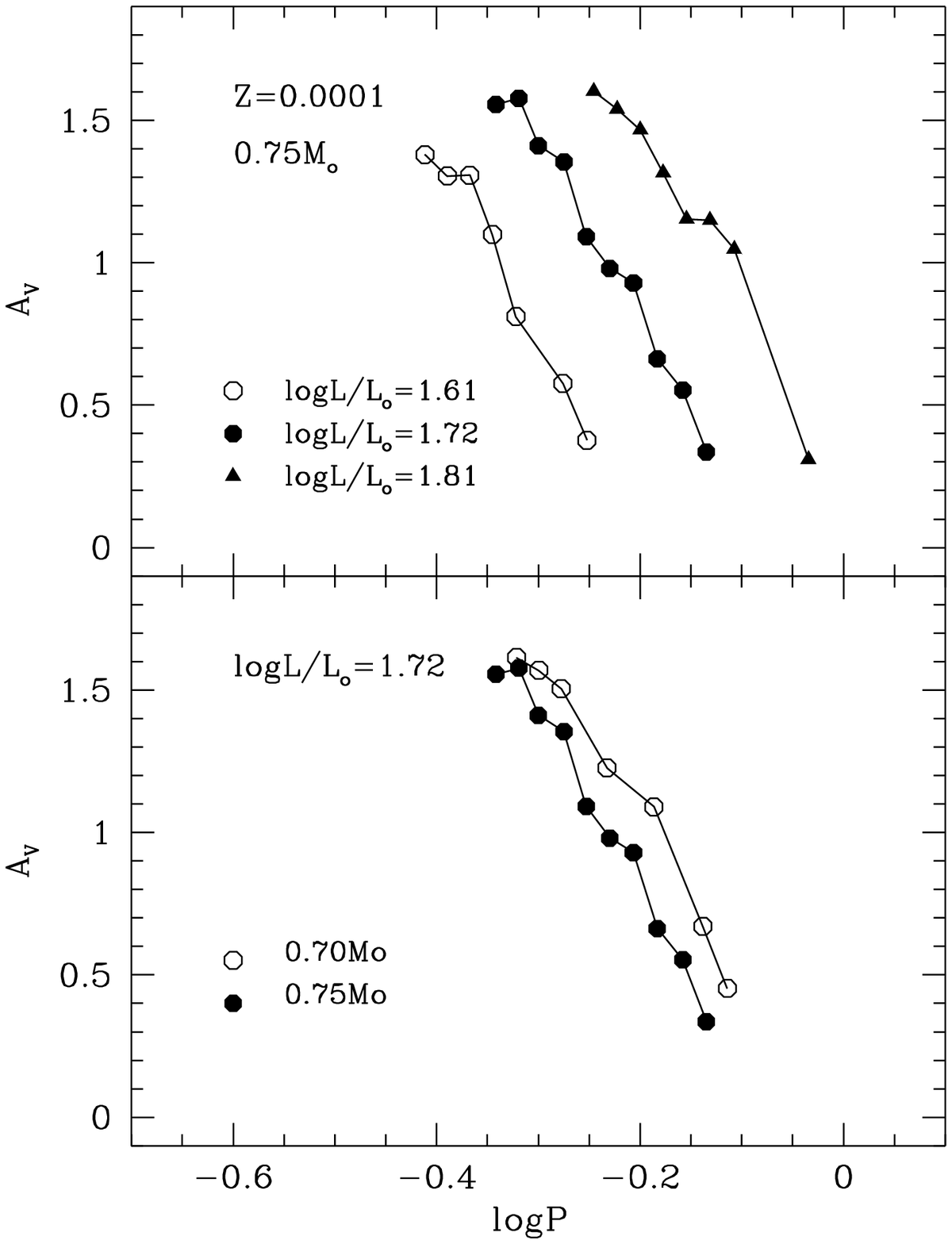}
\caption{Predicted Bailey diagram, V amplitude vs $\log P$, for fundamental 
pulsators constructed assuming fixed stellar mass and metallicity (top 
panel) as well as fixed stellar luminosity and  metallicity (bottom panel). 
Models from Di Criscienzo, Marconi \& Caputo (2002).}
\end{figure}

%Fig 7
\clearpage
\begin{figure}
\plotone{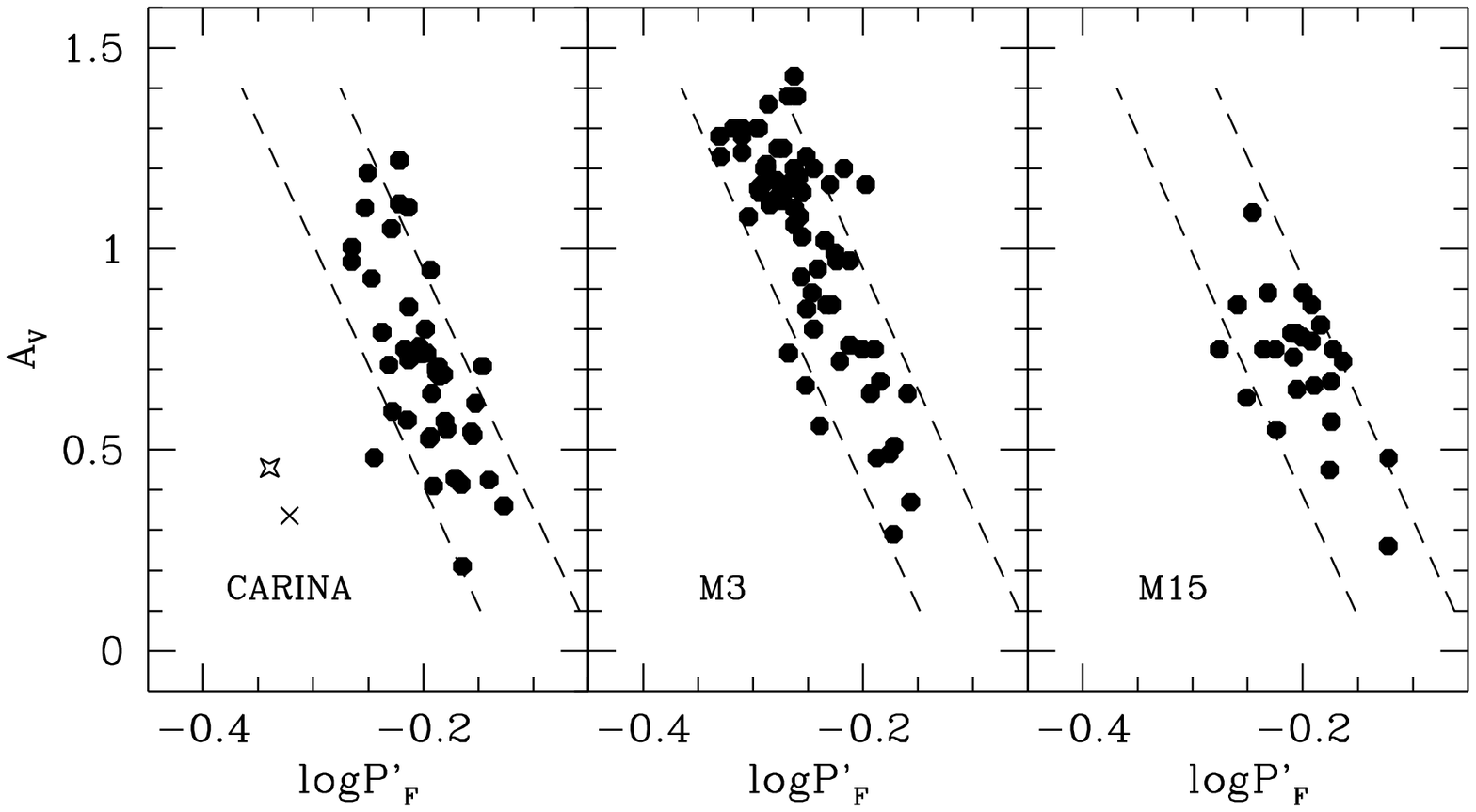}
\caption{Left: `Reduced' Bailey diagram $A_V$ vs log$P_F'$ for Carina 
$RRab$ variables (see text for details). The dashed lines trace in this 
plane a `reduced' period dispersion at constant amplitude, namely 
$\Delta$log$P_F'=0.09$. They were estimated using eq. (2) and by assuming 
a mass range of $0.65-0.80 M_\odot$. The middle and the right panel show 
the same data but for M3 and M15, respectively. Special symbols mark the 
variables, V158 (star) and V182 (cross), that show significantly shorter 
reduced periods. According to eq. (2), they should have masses $\approx 50\%$ 
larger than the bulk of the Carina $RRab$ variables.} 
\end{figure}

%Fig 8
\clearpage
\begin{figure}
\plotone{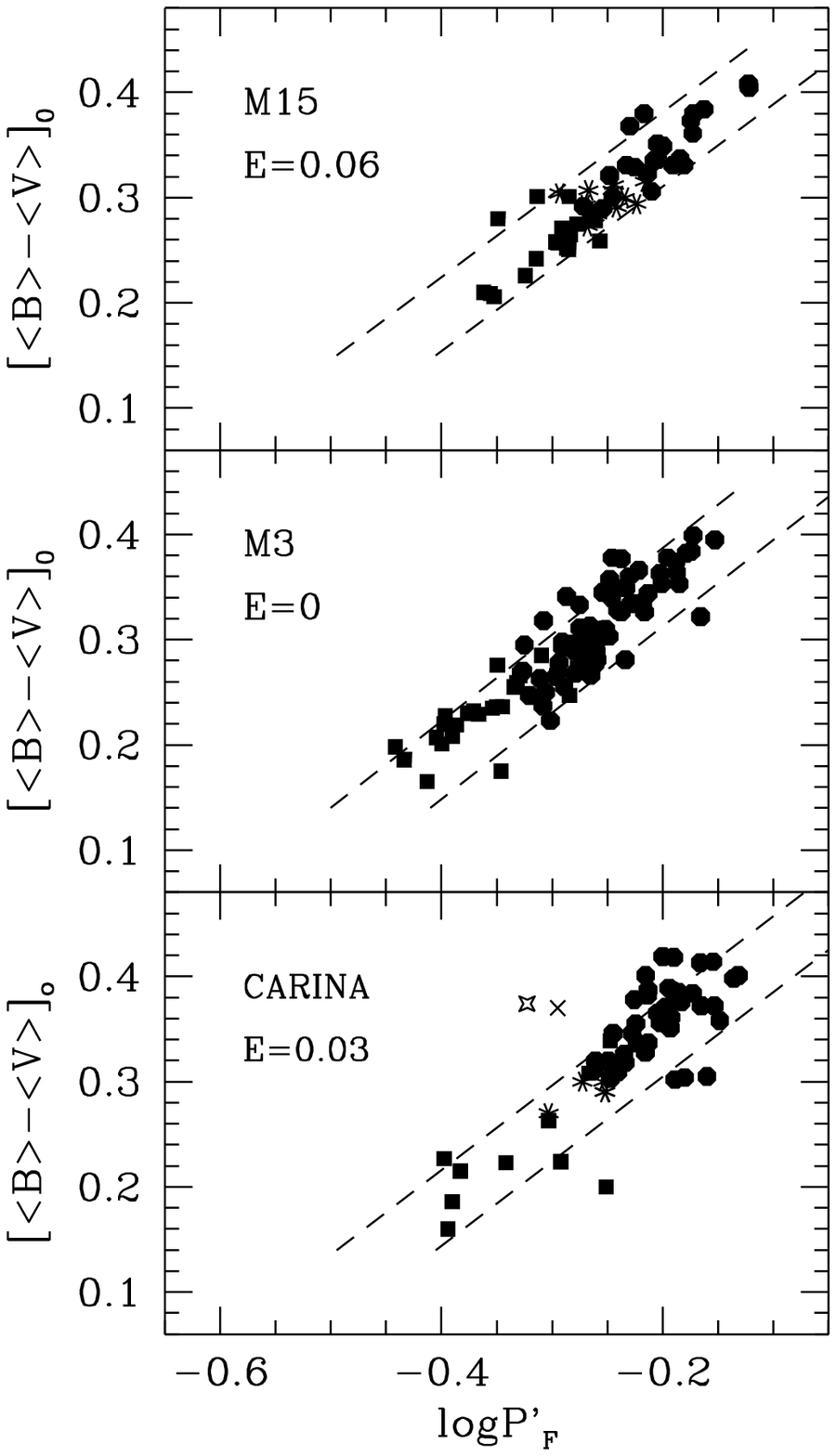}
\caption{Dereddened colors vs. reduced periods for RR Lyrae stars in 
M15, M3, and Carina. The dashed lines depict
the predicted period dispersion at constant color ($\Delta$log$P'_F=0.09$),
estimated assuming stellar masses ranging from 0.65 to 0.80 $M_\odot$. 
The present plot confirms that variables V158 (star) and V182 (cross) are 
significantly more massive than the bulk of Carina RR Lyrae stars 
(see also Fig. 7).}
\end{figure}

%Fig 9
\clearpage
\begin{figure}
\plotone{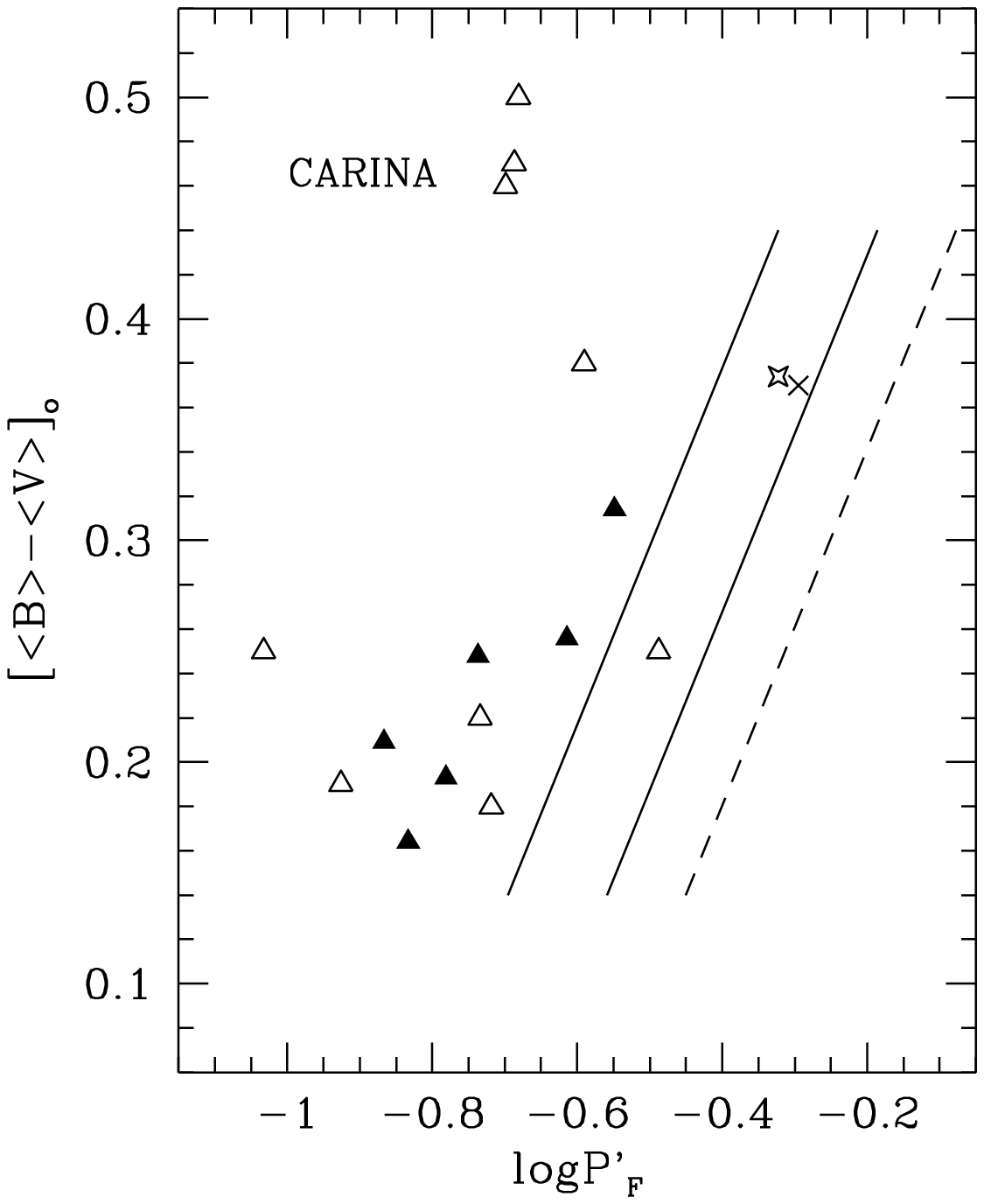}
\caption{Same as Fig. 8, but for variables brighter than $<V>=20$ mag. The
solid lines display the location in this plane of fundamental pulsators with 
masses 1.5 and 2 times (right to left) larger than the the bulk of RR Lyrae 
variables. The dashed line shows the mean locus of RR Lyrae stars. 
The distribution of Anomalous Cepheids show that these stars are massive 
pulsators. For comparison, the position of V158 and V182 are also displayed.}
\end{figure}

%Fig 10  
\clearpage
\begin{figure}
\plotone{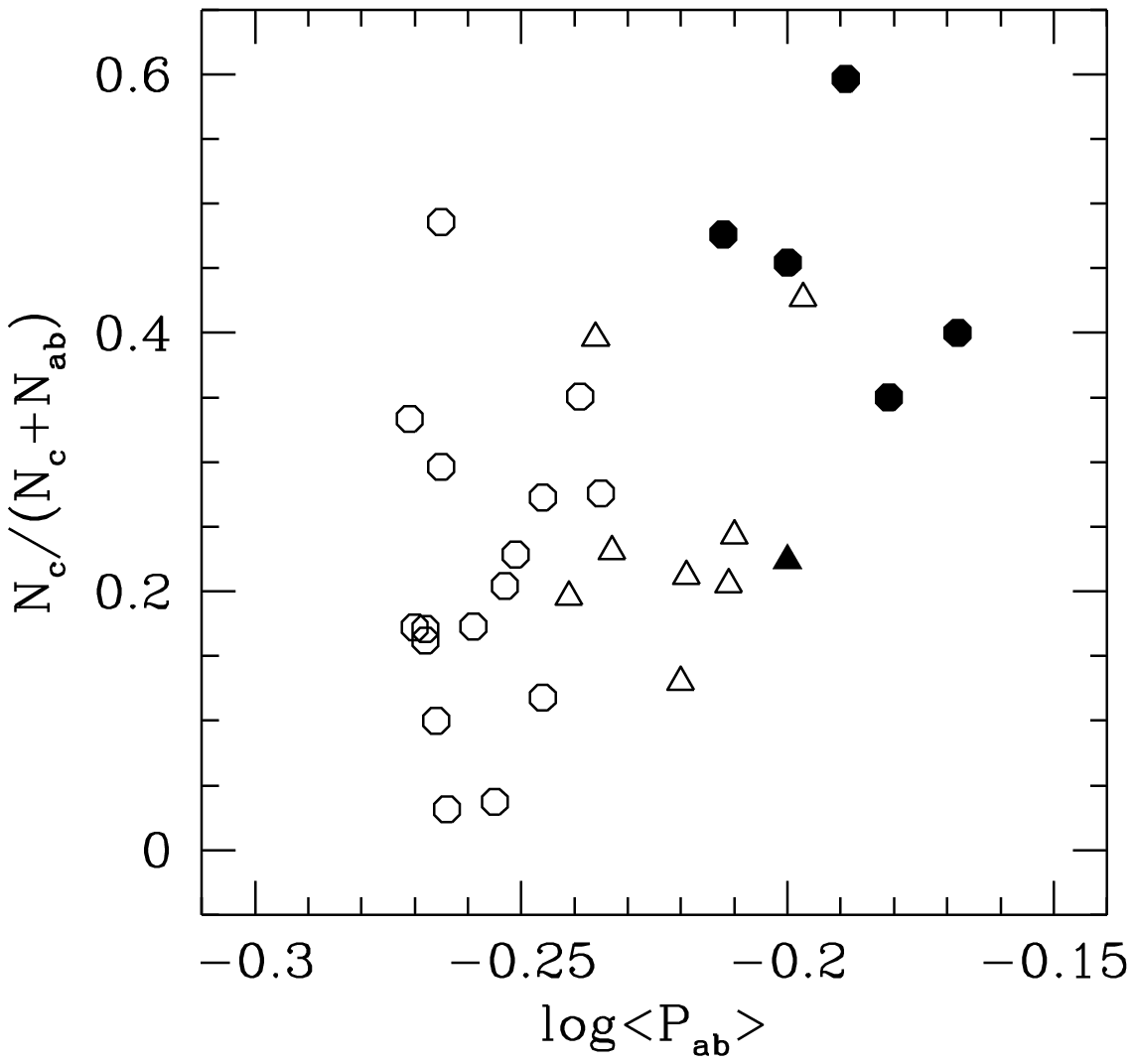}
\caption{Ratio between first overtone RR Lyrae ($RRc$) and the total number 
of RR Lyrae as a function of the mean period of fundamental ($RRab$) 
variables. Filled triangle refers to Carina, filled and open circles to 
RR Lyrae-rich OoII and OoI globular clusters, and open triangles to dSphs 
(see Table 4).}
\end{figure}

%Fig 11  
\clearpage
\begin{figure}
\plotone{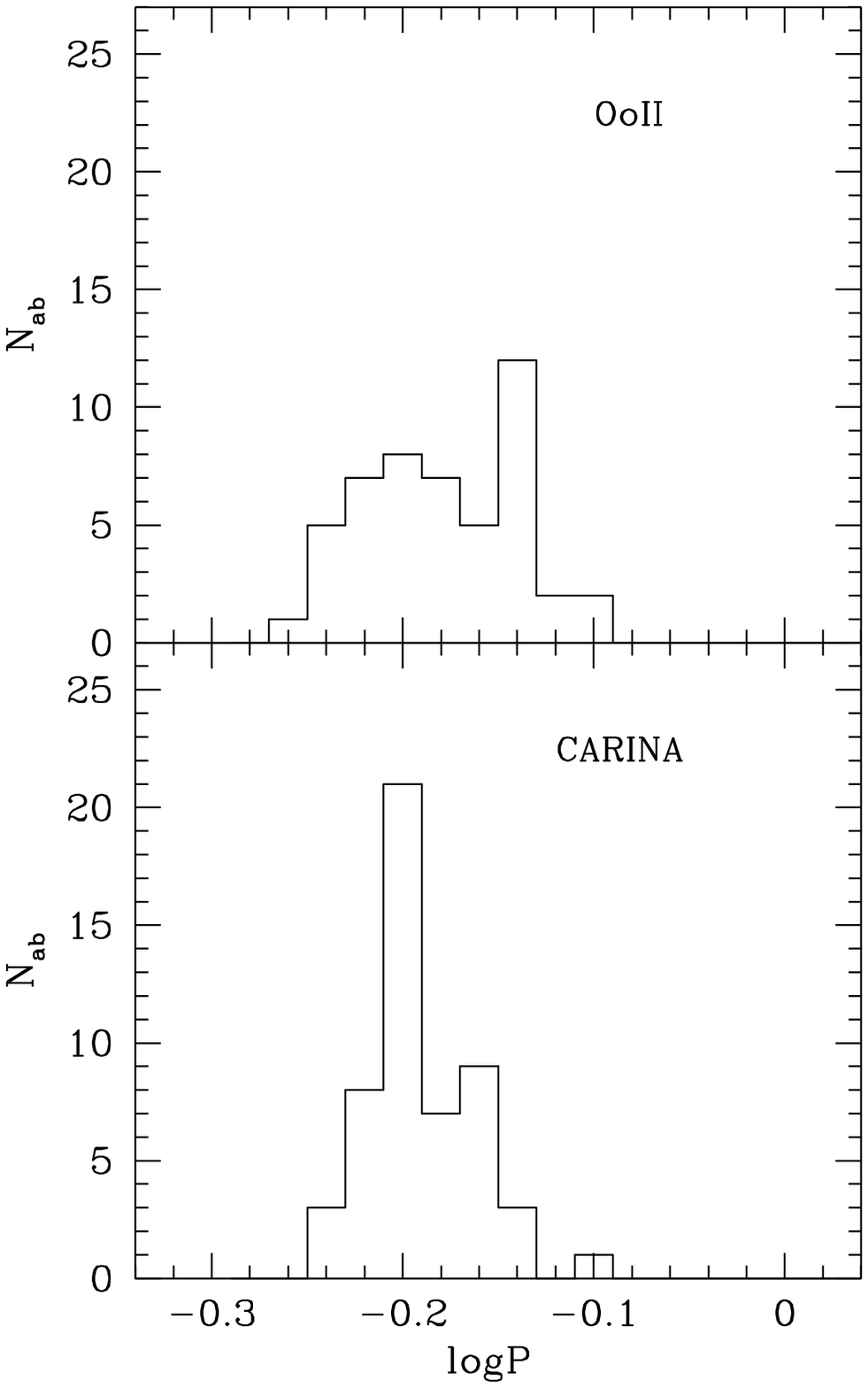}
\caption{Comparison between the period distribution of the Carina $RRab$ 
variables (bottom panel) with the cumulative distribution of $RRab$ in 
OoII Galactic globular clusters M15, M53, and M68 (top panel).}
\end{figure}

%Fig 12
\clearpage
\begin{figure}
\plotone{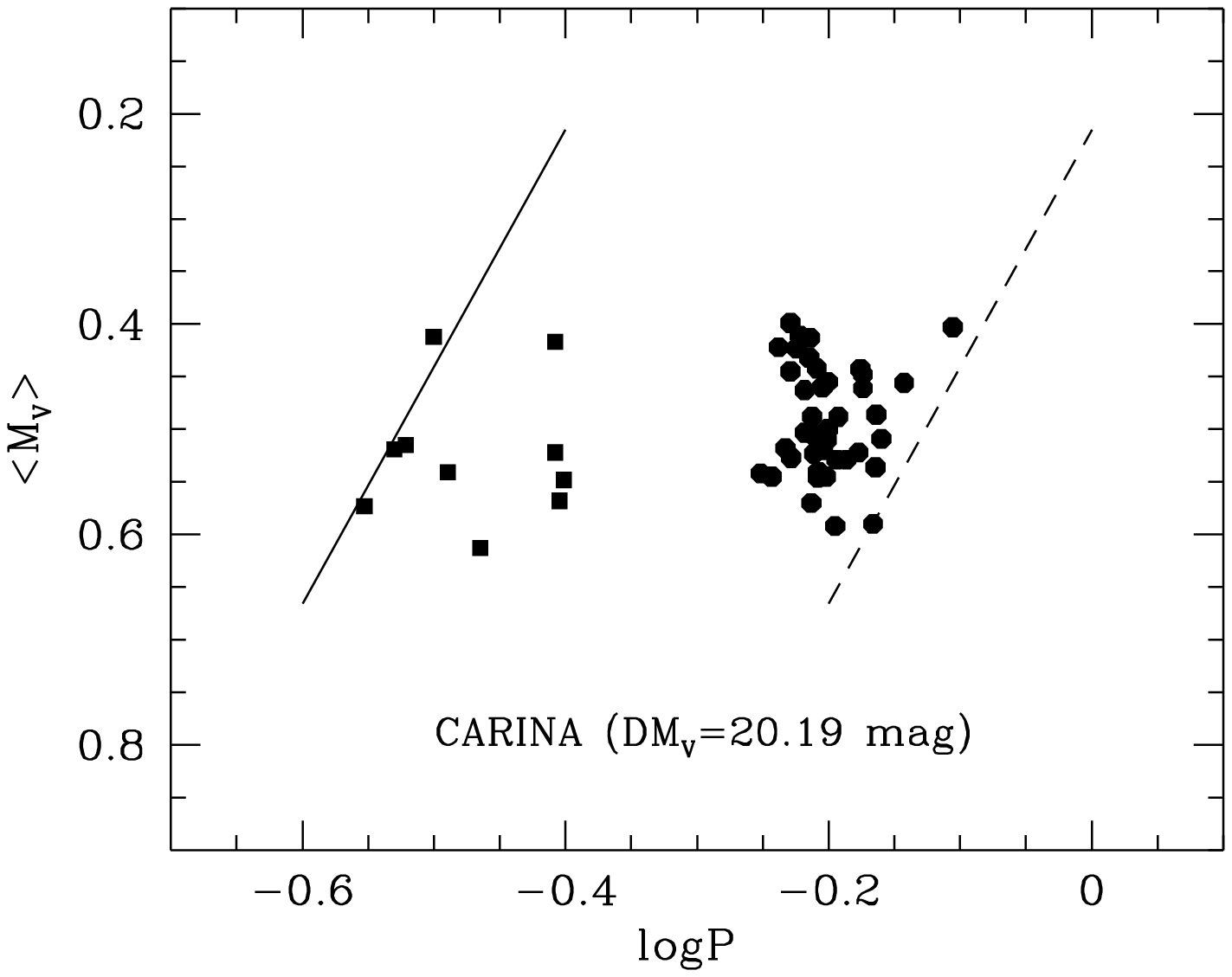}
\caption{Distribution of the Carina RR Lyrae stars in the $M_V$-log$P$ plane.
Superimposed are the predicted blue (solid line) and red (dashed line) edge 
of the instability strip. By matching observed and predicted blue edge 
(FOBE method, Caputo et al. 2000) we estimate an apparent distance modulus 
for Carina of $DM_V=20.19 \pm 0.04$ mag.}
\end{figure}

%Fig 13
\clearpage
\begin{figure}
%\plotone{dallora.fig13.ps}
\caption{Identification map for variable stars located inside chip \# 50.}
\end{figure}

%Fig 14
\begin{figure}
%\plotone{dallora.fig14.ps}
\caption{Same as Fig. 13, but for chip \# 51.}
\end{figure}

%Fig 15
\begin{figure}
%\plotone{dallora.fig15.ps}
\caption{Same as Fig. 13, but for chip \# 52.}
\end{figure}

%Fig 16
\begin{figure}
%\plotone{dallora.fig16.ps}
\caption{Same as Fig. 13, but for chip \# 53.}
\end{figure}

%Fig 17
\begin{figure}
%\plotone{dallora.fig17.ps}
\caption{Same as Fig. 13, but for chip \# 54.}
\end{figure}

%Fig 18
\begin{figure}
%\plotone{dallora.fig18.ps}
\caption{Same as Fig. 13, but for chip \# 55.}
\end{figure}

%Fig 19
\begin{figure}
%\plotone{dallora.fig19.ps}
\caption{Same as Fig. 13, but for chip \# 56.}
\end{figure}

%Fig 20  
\begin{figure}
%\plotone{dallora.fig20.ps}
\caption{Same as Fig. 13, but for chip \# 57.}
\end{figure}

\end{document}